\documentclass[12pt]{iopart}
\usepackage{graphicx}

\newcommand\lapp{\mathrel{\rlap{\lower4pt\hbox{\hskip1pt$\sim$}}
    \raise1pt\hbox{$<$}}}
\newcommand\gapp{\mathrel{\rlap{\lower4pt\hbox{\hskip1pt$\sim$}}
    \raise1pt\hbox{$>$}}}
\newcommand\eapp{\mathrel{\rlap{\raise2pt\hbox{\hskip0pt$\sim$}}
    \lower1pt\hbox{$-$}}}

\begin{document}

\title[The coincidence problem in linear dark energy models]
{The coincidence problem in linear dark energy models}

\author{P P Avelino\dag}

\address{\dag\ Centro de F\'{\i}sica do Porto e Departamento de F\'{\i}sica 
da Faculdade de Ci\^encias da Universidade do Porto, Rua do Campo Alegre 687,
4169-007, Porto, Portugal}

\begin{abstract}
We show that a solution to the the coincidence problem can be found in the 
context of a generic class of dark energy models with a scalar field, $\phi$, 
with a linear effective potential $V(\phi)$. We determine the fraction, $f$,  
of the total lifetime of the universe, $t_U$, which lies within the interval 
$[t_0-\Delta t_A,t_0+ \Delta t_A]$, where $t_0$ is the age of the universe 
at the present time, $\Delta t_A \equiv t_0-t_A$ and $t_A$ is the age of the 
universe when it starts to accelerate. We find 
that if we require $f$ to be larger than $0.1$ ($0.01$) then 
$1+\omega_{\phi0} \gapp 2 \times 10^{-2}$ ($1 \times 10^{-3}$), 
where $\omega_{\phi} \equiv p_\phi/\rho_\phi$. 
These results depend mainly on the linearity of the 
scalar field potential for $-V(\phi_0) \lapp V(\phi) \lapp V(\phi_0)$ and 
are weakly dependent on the specific form of $V(\phi)$ outside this range. 
We also show that if $\omega_{\phi0}$ is close to $-1$ then 
$\omega_{\phi0}+1 \sim 1.6\,({\tilde \omega}_\phi +1)$, 
where ${\tilde \omega}_\phi$ is 
the weighted average value of $\omega_{\phi}$ in the time interval 
$[0,t_0]$. We independently confirm current observational 
constraints on this class of models which give $\omega_{\phi0} \lapp -0.6$ 
and $t_U \gapp 2.4 \, t_0$ at the $2 \sigma$ level.

\end{abstract}

\pacs{98.80.-k, 98.80.Es, 95.35.+d, 12.60.-i}
\ead{ppavelin@fc.up.pt}
\maketitle

\section{Introduction}

In the last few years there has been a growing body of observational 
evidence strongly suggesting that we live in a (nearly) flat Universe 
which has recently entered an accelerating phase \cite{Perlmutter,Riess,
Tonry,wmap1,wmap}. If this acceleration is due to the presence of a tiny 
cosmological constant then we are living a very special phase of the 
Universe in which $\Omega_\Lambda \sim \Omega_m$. At earlier times 
$\Omega_m \sim 1$ ($\Omega_\Lambda \sim 0$) while at later times 
$\Omega_\Lambda \sim 1$ ($\Omega_m \sim 0$). This is known as the 
coincidence problem which asks whether this is just a coincidence 
or if there is a deeper explanation for such a fact. Of course there 
are alternative explanations for the recent acceleration of the 
Universe other than the cosmological constant. In the context of 
general relativity such a period of accelerated expansion must be 
induced by an exotic `dark energy' component violating the strong energy 
condition \cite{lambda,picon,Wang,solid,Bagla}, though this is not
necessarily so in the context of more general models 
(see for example \cite{brane}). 

There have been several attempts to explain this apparent coincidence in 
particular in the context of quintessence 
models. In some of the proposed models \cite{picon, malquarti} the dark 
energy density 
evolves from tracking behaviour in the radiation era towards a constant 
dark energy density in the matter era with the onset of acceleration being 
associated to the transition between the radiation and matter 
dominated epochs. Other attempts to solve the coincidence problem include 
models with alternate periods of matter and dark energy domination 
\cite{dodelson, griest}
or those where matter and quintessence fields are coupled in such a 
way that a nearly constant ratio between dark matter and dark energy densities 
is obtained at late times \cite{chimento1,chimento2,huey}. 
However all such attempts are only partially satisfactory since they 
do not in general explain why we are so close to the start of the 
first (there may be more than one) accelerating era, ignoring of course 
possible inflationary epoch(s) in the very early Universe. 

Another, more satisfactory, explanation for the apparent coincidence of 
dark matter and dark energy energy densities is found in the context of models in which 
the total universe lifetime is not much larger than the age of 
the universe today. A particular class of such models 
was recently studied in \cite{scherrer} (see also \cite{cai}) 
in the context of 
phantom dark energy scenarios \cite{caldwell1,caldwell2,nesseris}. 
The author  
has found that typically the fraction of the total lifetime of the universe for 
which the dark energy and dark matter densities are comparable is 
significant thus helping to solve the coincidence problem. However, 
the physical significance of these results is unclear since phantom models 
are expected to develop instabilities at the quantum level 
\cite{carroll,brunier,nojiri}.

In this paper we study a similar problem in the context of a cosmological 
model where a scalar field with a linear effective potential is the dark 
energy. Observational bounds on this type of models have 
been investigated in refs. \cite{wang,kallosh} and \cite{bek} (in the latter 
case in the context of varying alpha models). In this paper we shall independently 
confirm these constraints and determine the conditions that have to be
satisfied for the coincidence problem to be solved in this class of models.
In Sec. II we describe the linear model for the dark energy in detail  
and then discuss the results in Sec. III. Finally we summarize our 
results and briefly discuss future prospects in Sec. IV.

\section{The dynamics of the universe in the linear model}

We consider the dynamics of a flat homogeneous and isotropic 
Friedmann-Robertson-Walker 
(FRW) universe filled with matter and a scalar field $\phi$ which is fully 
described by 
\begin{eqnarray}
\label{eq1}
\frac{\ddot a}{a}=-H_i^2\left[\frac{\Omega_{mi}}{2} a^{-3} 
+  \frac{\Omega_\phi^*}{2} (1+3w_\phi)\right]\, , \\
\label{eq2}
{\ddot \phi}+3H{\dot \phi}=-\frac{dV}{d \phi} \, ,
\end{eqnarray}
where
\begin{equation} 
\Omega_\phi^* \equiv \Omega_\phi \frac{H^2}{H^2_i} = 
\frac{{\dot \phi}^2/2+V(\phi)}{H_i^2}\, , 
\label{eq3}
\end{equation} 
and  
\begin{equation} 
\omega_\phi=\frac{{\dot \phi}^2/2-V(\phi)}{{\dot \phi}^2/2+V(\phi)}\, . 
\label{eq4}
\end{equation} 
Here $a$ is the scale factor, $H\equiv {\dot a}/a$, a dot represents a 
derivative with respect to physical time, $\Omega_\phi=\rho_\phi/\rho_c$, 
$\Omega_m=\rho_m/\rho_c$, $\rho_c$ is the critical density, 
the subscript `$i$' means that the variables are to be evaluated at 
some initial time $t_i \ll t_0$ deep into the matter era (so that 
$H_i=2/(3t_i)$), $t_0$ is the age of the universe at the present time, 
we took $a_i=1$ and we are using units in which $8\pi G/3 =1$. We also assume 
that the kinetic energy of the 
field $\phi$ at the initial time is completely determined by the scalar 
field potential, that is ${\dot \phi}_i \equiv {\dot \phi}(t_i)$ has 
no memory of initial conditions. Given that we are taking the initial time, $t_i$, to 
be deep into the matter era this means that \cite{bek}
\begin{equation} 
{\dot \phi}_i \equiv {\dot \phi}(t_i) = -\frac{2}{9 H_i} 
\frac{dV}{d \phi} \, . 
\label{eq5}
\end{equation} 
Throughout this paper we will assume that $V(\phi)$  
is a linear function of $\phi$, namely 
\begin{equation} 
V(\phi) = V_0 + \frac{dV}{d\phi} \left(\phi-\phi_0\right)\, , 
\label{eq6}
\end{equation} 
where ${dV}/{d\phi}$ is assumed to be a negative constant and the subscript 
`$0$' means that the variables are to be evaluated at the present time $t_0$. Given that 
observations constrain $\omega_{\phi0}$ to be close to $-1$ in this 
paper we will only consider models for which
\begin{equation} 
\omega_{\phi 0} \sim -1 + \frac{{\dot \phi}^2_0}{V_0}\, , 
\label{eq7}
\end{equation}
is a good approximation. In fact, since $\omega_\phi \to -1$ very rapidly as 
we move backwards in time \cite{kallosh,wang,bek}, eqn.(\ref{eq7}) 
is also a good approximation for $t < t_0$. For 
simplicity, we shall also assume that $\Omega_{m0}=0.3$ ($\Omega_{\phi0}=0.7$). 
We have checked that our results are weakly dependent on this assumption 
as long as we consider $\Omega_{m0}$ to be within the range allowed by 
current observational constraints.

Hence, for $\omega_{\phi0} \sim -1$ the cosmological evolution of $\phi$ up to the 
present time is such that ${\dot \phi}$ is completely determined up to 
a normalization factor proportional to $dV/d\phi$. Consequently, from eqn. 
(\ref{eq7}) we have 
\begin{equation} 
\omega_\phi + 1 \propto \left(\frac{dV}{d\phi}\right)^2\, . 
\label{eq8}
\end{equation}
for $t \le t_0$. The weighted average value of $\omega_\phi$ in the time 
interval $[0,t_0]$ can be 
calculated for this type of models and related to $\omega_{\phi 0}$. 
We have found that 
\begin{equation} 
\omega_{\phi0}+1 \sim 1.6\,({\tilde \omega}_\phi +1)\, , 
\label{eq8a}
\end{equation}
where
\begin{equation} 
{\tilde \omega}_\phi \equiv 
\frac{\int_0^{a_0} da \, \Omega_\phi \ \omega_\phi}{\int_0^{a_0} da \,  
\Omega_\phi}. 
\label{eq8b}
\end{equation}
Current observational bounds on the value of ${\tilde \omega}_\phi$ which 
give ${\tilde \omega}_\phi \lapp -0.8$ at the $2$ sigma level 
\cite{wmap,caldwell} can be easily 
translated into $\omega_{\phi 0} \lapp -0.6$ and $t_U \gapp 2.4 \, t_0$ 
consistent with the results of ref. \cite{wang}.

\begin{figure}
\begin{center}
\includegraphics*[width=9cm]{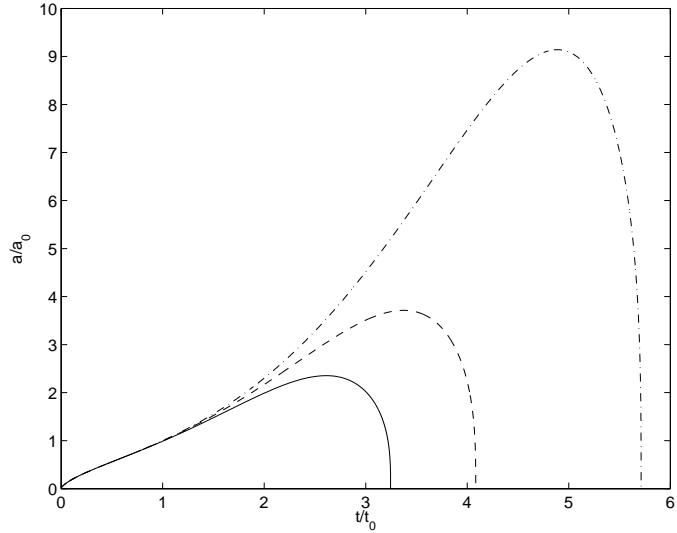}
\end{center}
\caption{\label{fig1} The evolution of the scale factor as a 
function of physical time for $\omega_{\phi 0}=-0.8,-0.9,-0.95$ 
(solid line, dashed line and dot-dashed lines respectively). The closer 
$\omega_{\phi 0}$ is to $-1$ the larger is the total lifetime of the universe.
}
\end{figure}

\section{Results and discussion}

In Fig.~\ref{fig1} we plot the evolution of the scale factor (in units 
of $a_0$) as a function of physical time (in units of $t_0$) for 
$\omega_{\phi 0}=-0.8,-0.9,-0.95$ (solid line, dashed line and 
dot-dashed lines respectively). Note that since all the models 
considered have $\omega_{\phi 0} \sim -1$ the age of the universe at the 
present time is approximately the same for all the $3$ models considered. 
We see that all the models have a matter dominated phase with 
$a \propto t^{2/3}$ followed by an accelerating phase with ${\ddot a} > 0$ 
and then by a rapid collapse. It is also obvious from Fig.~\ref{fig1} that 
the closer $\omega_{\phi 0}$ is to $-1$ the longer is duration of 
accelerating phase and hence the larger is the total lifetime of the universe.

\begin{figure}
\begin{center}
\includegraphics*[width=9cm]{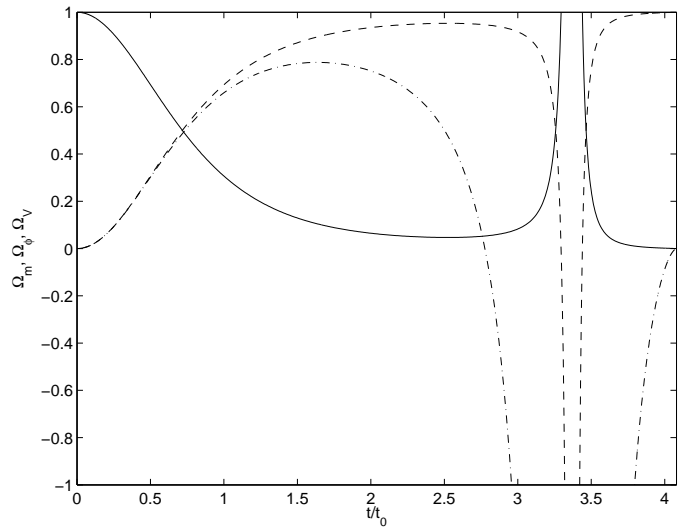}
\end{center}
\caption{\label{fig2} The evolution of $\Omega_m$, $\Omega_\phi$ and $\Omega_V 
\equiv V(\phi)/H^2$ (solid, dashed and dot-dashed lines respectively) as a 
function of cosmic time for the model with $\omega_{\phi 0}=-0.9$}
\end{figure}

In  Fig.~\ref{fig2} we plot the evolution of $\Omega_m$, $\Omega_\phi$ and 
$\Omega_V \equiv V(\phi)/H^2$ (solid, dashed and dot-dashed lines 
respectively) as a function of cosmic time for the model with 
$\omega_{\phi 0}=-0.9$. Again we see 
that the universe has $\Omega_m \sim 1$ at early times and then starts 
to accelerate near the present time. In the accelerating phase the evolution 
of $\phi$ is slow and the main contribution to the energy density of the 
universe comes from $V(\phi)$. During most of this phase 
$\Omega \sim \Omega_V \sim 1$. At some point the energy density of the 
scalar field $\phi$ becomes small enough and the dynamics of the universe is 
again dominated by matter and the universe starts decelerating again. 
Later on $V(\phi)$ turns negative and the universe starts collapsing very 
rapidly with the energy density being dominated by the kinetic energy 
density of the scalar field $\phi$. In this phase ${\dot \phi} \propto a^{-3}$ 
and $a \propto (t_U-t)^{1/3}$, where $t_U$ is the total lifetime of the 
universe.

\begin{figure}
\begin{center}
\includegraphics*[width=9cm]{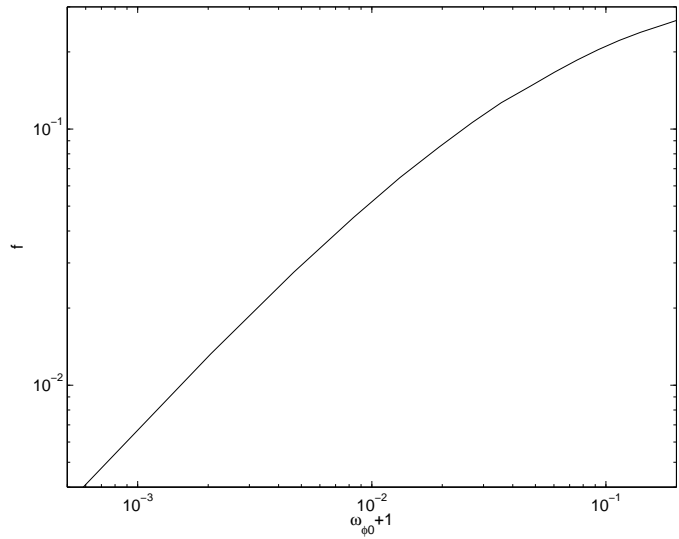}
\end{center}
\caption{\label{fig3} The fraction, $f$, of the total lifetime of the universe 
which lies within the interval $[t_0-t_A, t_0+t_A]$ as a function of 
$\omega_{\phi0}$. Here $t_0$ and $t_A$ are respectively the age of the 
universe at the present time and when it starts to accelerate. 
Note that $f$ is 
a decreasing function of $\omega_{\phi0}$ and that for $\omega_{\phi0}$ close to 
$-1$ there is an almost linear relation between $f$ and $\omega_{\phi0}+1$.}
\end{figure}

In Fig.~\ref{fig3} we plot the fraction, $f$, of the total lifetime of the 
universe which lies within the interval $[t_0-\Delta t_A, t_0+\Delta t_A]$ 
where $t_0$ is the age of the Universe at the present time, 
$\Delta t_A \equiv t_0-t_A$ and $t_A$ is 
age of the universe when it starts to accelerate. We see that $f$ is a 
increasing function of 
$\omega_{\phi0}+1$ and that for $\omega_{\phi0}$ close to $-1$ there is an 
almost linear relation between $f$ and $\omega_{\phi0}+1$. This linear relation 
is to be expected since for $\omega_{\phi0}$ very close to $-1$ most 
of the lifetime of the universe is spent in the accelerating phase. 
During most of the accelerating phase $V \sim V_0$ so that 
we may estimate the variation of the value of $\phi$ by
\begin{equation} 
\Delta \phi \sim \frac{V_0}{dV/d\phi}\, .  
\label{eq9}
\end{equation}
However, in this phase the scalar field $\phi$ is slowly rolling down 
the scalar field potential with $3H {\dot \phi} \sim -dV/d\phi$. Hence, we 
have that 
\begin{equation} 
\Delta \phi \propto -\frac{dV}{d\phi} \Delta \ln a \propto -\frac{dV}{d\phi} 
\Delta t \, , 
\label{eq10}
\end{equation}
where $\Delta t$ is the duration of the accelerating phase.
From eqns. (\ref{eq9}) and (\ref{eq10}) we see that 
\begin{equation} 
\Delta t \propto \left(\frac{dV}{d\phi}\right)^{-2} \propto 
\left(\omega_\phi+1\right)^{-1} \, ,  
\label{eq11}
\end{equation}
so that $f \equiv 2 (t_0-t_A)/t_U \propto \omega_\phi+1$ for large enough $t_U$ 
which is confirmed by looking at Fig.~\ref{fig3} (note that for $t_U$ sufficiently 
large $\Delta t \sim t_U$). We note that the value 
of $f$ for which the observed value of $t_A$ could be considered a 
coincidence is of course not well defined. However, from Fig.~\ref{fig3} 
we see that if we require $f$ to be larger than $0.1$ ($0.01$) then 
$1+\omega_{\phi0} \gapp 2 \times 10^{-2}$ ($1 \times 10^{-3}$).

\section{Conclusions}

In this paper we have investigated the coincidence problem in the 
context of a generic class of linear dark energy models. 
If we require our model to satisfy current observational constraints 
providing at the same time a solution to the coincidence problem then
$ 2 \times 10^{-2} (1 \times 10^{-3}) \lapp 1+\omega_{\phi0} \lapp 0.4$. 
Of course the lower limit depends on how conservative our 
criterion is. We again emphasize that our results 
depend mainly on the linearity of the scalar field potential for 
$-V(\phi_0) \lapp V(\phi) \lapp V(\phi_0)$ and are weakly dependent 
on the specific form of $V(\phi)$ outside this range. 
It is very interesting that a straightforward solution to the coincidence 
problem in the simplest generalization 
of the standard cosmological constant scenario would require a significant 
departure of $\omega_{\phi0}$ from $-1$ which may eventually be measured 
by the next generation of cosmological observations \cite{kallosh}. 

\ack
I am grateful to Margarida Cunha, Carlos Martins, Joana Oliveira and Pedro Viana for useful 
discussions. This work was partially funded by Funda{\c c}\~ao para a Ci\^encia 
e Tecnologia (Portugal) under contract POCTI/FP/FNU/50161/2003.


\section*{References}
\begin{thebibliography}{30}
\expandafter\ifx\csname natexlab\endcsname\relax\def\natexlab#1{#1}\fi
\expandafter\ifx\csname bibnamefont\endcsname\relax
  \def\bibnamefont#1{#1}\fi
\expandafter\ifx\csname bibfnamefont\endcsname\relax
  \def\bibfnamefont#1{#1}\fi
\expandafter\ifx\csname citenamefont\endcsname\relax
  \def\citenamefont#1{#1}\fi
\expandafter\ifx\csname url\endcsname\relax
  \def\url#1{\texttt{#1}}\fi
\expandafter\ifx\csname urlprefix\endcsname\relax\def\urlprefix{URL }\fi
\providecommand{\bibinfo}[2]{#2}
\providecommand{\eprint}[2][]{\url{#2}}

\bibitem{Perlmutter}
\bibinfo{author}{\bibfnamefont{S.}~\bibnamefont{Perlmutter}}
  \bibnamefont{et~al.}, \bibinfo{journal}{Astrophys. J.}
  \textbf{\bibinfo{volume}{517}}, \bibinfo{pages}{565} (\bibinfo{year}{1999}),
  \eprint{astro-ph/9812133}.

\bibitem{Riess}
\bibinfo{author}{\bibfnamefont{A.~G.} \bibnamefont{Riess}}
  \bibnamefont{et~al.}, \bibinfo{journal}{Astrophys. J.}
  \textbf{\bibinfo{volume}{560}}, \bibinfo{pages}{49} (\bibinfo{year}{2001}),
  \eprint{astro-ph/0104455}.

\bibitem{Tonry}
\bibinfo{author}{\bibfnamefont{J.}~\bibnamefont{Tonry}} 
  \bibnamefont{et~al.}, \bibinfo{journal}{Astrophys. J.}
  \textbf{\bibinfo{volume}{594}}, \bibinfo{pages}{1} 
  (\bibinfo{year}{2003}), \eprint{astro-ph/0305008}.

\bibitem{wmap1}
\bibinfo{author}{\bibfnamefont{C.~L.} \bibnamefont{Bennett}}
  \bibnamefont{et~al.} \bibinfo{journal}{Astrophys. J. Suppl.}
  \textbf{\bibinfo{volume}{148}}, \bibinfo{pages}{1} 
  (\bibinfo{year}{2003}), \eprint{astro-ph/0302207}.

\bibitem{wmap}
\bibinfo{author}{\bibfnamefont{D.~N.} \bibnamefont{Spergel}}
  \bibnamefont{et~al.} \bibinfo{journal}{Astrophys. J. Suppl.}
  \textbf{\bibinfo{volume}{148}}, \bibinfo{pages}{175} 
  (\bibinfo{year}{2003}), \eprint{astro-ph/0302209}.

\bibitem{lambda}
\bibinfo{author}{\bibfnamefont{S.~M.} \bibnamefont{Carroll}},
  \bibinfo{journal}{Living Rev. Rel.} \textbf{\bibinfo{volume}{4}},
  \bibinfo{pages}{1} (\bibinfo{year}{2001}), \eprint{astro-ph/0004075}.

\bibitem{picon}
\bibinfo{author}{\bibfnamefont{C.}~\bibnamefont{Armendariz-Picon}},
  \bibinfo{author}{\bibfnamefont{V.}~\bibnamefont{Mukhanov}} \bibnamefont{and}
  \bibinfo{author}{\bibfnamefont{P.~J.} \bibnamefont{Steinhardt}},
  \bibinfo{journal}{Phys. Rev.} \textbf{\bibinfo{volume}{D63}},
  \bibinfo{pages}{103510} (\bibinfo{year}{2001}), \eprint{astro-ph/0006373}.

\bibitem{Wang}
\bibinfo{author}{\bibfnamefont{L.}~\bibnamefont{Wang}},
  \bibinfo{author}{\bibfnamefont{R.}~\bibnamefont{Caldwell}},
  \bibinfo{author}{\bibfnamefont{J.}~\bibnamefont{Ostriker}} \bibnamefont{and}
  \bibinfo{author}{\bibfnamefont{P.}~\bibnamefont{Steinhardt}},
  \bibinfo{journal}{Astrophys. J.} \textbf{\bibinfo{volume}{530}},
  \bibinfo{pages}{17} (\bibinfo{year}{2000}), \eprint{astro-ph/9901388}.

\bibitem{solid}
\bibinfo{author}{\bibfnamefont{M.}~\bibnamefont{Bucher}} \bibnamefont{and}
  \bibinfo{author}{\bibfnamefont{D.~N.} \bibnamefont{Spergel}},
  \bibinfo{journal}{Phys. Rev.} \textbf{\bibinfo{volume}{D60}},
  \bibinfo{pages}{043505} (\bibinfo{year}{1999}), \eprint{astro-ph/9812022}.

\bibitem{Bagla}
\bibinfo{author}{\bibfnamefont{J.}~\bibnamefont{Bagla}},
  \bibinfo{author}{\bibfnamefont{H.}~\bibnamefont{Jassal}} \bibnamefont{and}
  \bibinfo{author}{\bibfnamefont{T.}~\bibnamefont{Padmanabhan}},
  \bibinfo{journal}{Phys. Rev.} \textbf{\bibinfo{volume}{D67}},
  \bibinfo{pages}{063504} (\bibinfo{year}{2003}), \eprint{astro-ph/0212198}.

\bibitem{brane}
\bibinfo{author}{\bibfnamefont{P.~P.}~\bibnamefont{Avelino}} \bibnamefont{and}
  \bibinfo{author}{\bibfnamefont{C.~J.~A.~P.}~\bibnamefont{Martins}},
  \bibinfo{journal}{Astrophys. J.} \textbf{\bibinfo{volume}{565}},
  \bibinfo{pages}{661} (\bibinfo{year}{2002}), \eprint{astro-ph/0106274}.

\bibitem{malquarti}
\bibinfo{author}{\bibfnamefont{M.}~\bibnamefont{Malquarti}}, 
  \bibinfo{author}{\bibfnamefont{E.~J.}~\bibnamefont{Copeland}}, 
  \bibnamefont{and}
  \bibinfo{author}{\bibfnamefont{A.~R.}~\bibnamefont{Liddle}},
  \bibinfo{journal}{Phys. Rev.} \textbf{\bibinfo{volume}{D68}},
  \bibinfo{pages}{023512} (\bibinfo{year}{2003}), \eprint{astro-ph/0304277}.

\bibitem{dodelson}
\bibinfo{author}{\bibfnamefont{S.}~\bibnamefont{Dodelson}}, 
  \bibinfo{author}{\bibfnamefont{M.}~\bibnamefont{Kaplinghat}} 
  \bibnamefont{and}
  \bibinfo{author}{\bibfnamefont{E.}~\bibnamefont{Stewart}},
  \bibinfo{journal}{Phys. Rev. Lett.} \textbf{\bibinfo{volume}{85}},
  \bibinfo{pages}{5276} (\bibinfo{year}{2000}), \eprint{astro-ph/0002360}.

\bibitem{griest}
\bibinfo{author}{\bibfnamefont{K.}~\bibnamefont{Griest}},
  \bibinfo{journal}{Phys. Rev.} \textbf{\bibinfo{volume}{D66}},
  \bibinfo{pages}{123501} (\bibinfo{year}{2002}), \eprint{astro-ph/0202052}.

\bibitem{chimento1}
\bibinfo{author}{\bibfnamefont{L.~P.}~\bibnamefont{Chimento}}, 
  \bibinfo{author}{\bibfnamefont{A.~S.}~\bibnamefont{Jakubi}},
  \bibinfo{author}{\bibfnamefont{D.}~\bibnamefont{Pavon}},  
  \bibnamefont{and}
  \bibinfo{author}{\bibfnamefont{W.}~\bibnamefont{Zimdahl}},
  \bibinfo{journal}{Phys. Rev.} \textbf{\bibinfo{volume}{D67}},
  \bibinfo{pages}{083513} (\bibinfo{year}{2003}), \eprint{astro-ph/0303145}.

\bibitem{chimento2}
\bibinfo{author}{\bibfnamefont{L.~P.}~\bibnamefont{Chimento}}, 
  \bibinfo{author}{\bibfnamefont{A.~S.}~\bibnamefont{Jakubi}},  
  \bibnamefont{and}
  \bibinfo{author}{\bibfnamefont{D.}~\bibnamefont{Pavon}},
  \bibinfo{journal}{Phys. Rev.} \textbf{\bibinfo{volume}{D67}},
  \bibinfo{pages}{087302} (\bibinfo{year}{2003}), \eprint{astro-ph/0303160}.

\bibitem{huey}
\bibinfo{author}{\bibfnamefont{G.}~\bibnamefont{Huey}} \bibnamefont{and} 
  \bibinfo{author}{\bibfnamefont{B.}~\bibnamefont{Wandelt}}
  (\bibinfo{year}{2004}), \eprint{astro-ph/0407196}.

\bibitem{scherrer}
\bibinfo{author}{\bibfnamefont{R.~J.}~\bibnamefont{Scherrer}}, 
  (\bibinfo{year}{2004}), \eprint{astro-ph/0410508}.

\bibitem{cai}
\bibinfo{author}{\bibfnamefont{R.-G.}~\bibnamefont{Cai}}, 
  \bibnamefont{and}
  \bibinfo{author}{\bibfnamefont{A.}~\bibnamefont{Wang}},
  (\bibinfo{year}{2004}), \eprint{hep-th/0411025}.

\bibitem{caldwell1}
\bibinfo{author}{\bibfnamefont{R.~R.}~\bibnamefont{Caldwell}}, 
  \bibinfo{journal}{Phys. Lett.} \textbf{\bibinfo{volume}{B545}},
  \bibinfo{pages}{23} (\bibinfo{year}{2002}), \eprint{astro-ph/9908168}.

\bibitem{caldwell2}
\bibinfo{author}{\bibfnamefont{R.~R.}~\bibnamefont{Caldwell}}, 
  \bibinfo{author}{\bibfnamefont{N.~N.}~\bibnamefont{Weinberg}}, 
  \bibnamefont{and}
  \bibinfo{author}{\bibfnamefont{M.}~\bibnamefont{Kamionkowski}},
  \bibinfo{journal}{Phys. Rev. Lett.} \textbf{\bibinfo{volume}{91}},
  \bibinfo{pages}{071301} (\bibinfo{year}{2003}), \eprint{astro-ph/0302506}.

\bibitem{nesseris}
\bibinfo{author}{\bibfnamefont{S.}~\bibnamefont{Nesseris}}, 
  \bibnamefont{and}
  \bibinfo{author}{\bibfnamefont{L.}~\bibnamefont{Perivolaropoulos}},
  (\bibinfo{year}{2003}), \eprint{astro-ph/0410309}.

\bibitem{carroll}
\bibinfo{author}{\bibfnamefont{S.~M.}~\bibnamefont{Carroll}}, 
  \bibinfo{author}{\bibfnamefont{M.}~\bibnamefont{Hoffman}}, 
  \bibnamefont{and}
  \bibinfo{author}{\bibfnamefont{M.}~\bibnamefont{Trodden}},
  \bibinfo{journal}{Phys. Rev.} \textbf{\bibinfo{volume}{D68}},
  \bibinfo{pages}{023509} (\bibinfo{year}{2003}), \eprint{astro-ph/0301273}.

\bibitem{brunier}
\bibinfo{author}{\bibfnamefont{T.}~\bibnamefont{Brunier}}, 
  \bibinfo{author}{\bibfnamefont{V.~K.}~\bibnamefont{Onemli}},
  \bibnamefont{and}
  \bibinfo{author}{\bibfnamefont{R.~P.}~\bibnamefont{Woodard}},
  (\bibinfo{year}{2004}), \eprint{gr-qc/0408080}.

\bibitem{nojiri}
\bibinfo{author}{\bibfnamefont{S.}~\bibnamefont{Nojiri}}, 
  \bibnamefont{and}
  \bibinfo{author}{\bibfnamefont{S.~D.}~\bibnamefont{Odintsov}},
  (\bibinfo{year}{2004}), \eprint{hep-th/0408170}.

\bibitem{wang}
\bibinfo{author}{\bibfnamefont{Y.} \bibnamefont{Wang}}, 
  \bibinfo{author}{\bibfnamefont{J.~M.}~\bibnamefont{Kratochvil}},
  \bibinfo{author}{\bibfnamefont{A.}~\bibnamefont{Linde}} \bibnamefont{and}
  \bibinfo{author}{\bibfnamefont{M.}~\bibnamefont{Shmakova}}
  (\bibinfo{year}{2004}), \eprint{astro-ph/0409264}.

\bibitem{kallosh}
\bibinfo{author}{\bibfnamefont{R.}~\bibnamefont{Kallosh}}, 
  \bibinfo{author}{\bibfnamefont{J.~M.}~\bibnamefont{Kratochvil}},
  \bibinfo{author}{\bibfnamefont{A.}~\bibnamefont{Linde}},
  \bibinfo{author}{\bibfnamefont{E.~V.}~\bibnamefont{Linder}} \bibnamefont{and}
  \bibinfo{author}{\bibfnamefont{M.}~\bibnamefont{Shmakova}},
  \bibinfo{journal}{JCAP} \textbf{\bibinfo{volume}{0310}},
  \bibinfo{pages}{015} (\bibinfo{year}{2004}), \eprint{astro-ph/0307185}.

\bibitem{bek}
\bibinfo{author}{\bibfnamefont{P.~P.}~\bibnamefont{Avelino}}, 
  \bibinfo{author}{\bibfnamefont{C.~J.~A.~P.}~\bibnamefont{Martins}} 
  \bibnamefont{and}
  \bibinfo{author}{\bibfnamefont{J.~C.~R.~E}~\bibnamefont{Oliveira}}, 
  \bibinfo{journal}{Phys. Rev.} \textbf{\bibinfo{volume}{D70}},
  \bibinfo{pages}{083506} (\bibinfo{year}{2004}), \eprint{astro-ph/0402379}.

\bibitem{caldwell}
\bibinfo{author}{\bibfnamefont{R.~R.}~\bibnamefont{Caldwell}}, 
  \bibnamefont{and}
  \bibinfo{author}{\bibfnamefont{M.}~\bibnamefont{Doran}},
  \bibinfo{journal}{Phys. Rev.} \textbf{\bibinfo{volume}{D69}},
  \bibinfo{pages}{103517} (\bibinfo{year}{2004}), \eprint{astro-ph/0305334}.

\end{thebibliography}
\end{document}